\documentstyle[12pt]{article}

\begin{document}

\title{Planary Symmetric Static Worlds with Massless Scalar Sources}
\author{{Ciprian Dariescu\thanks{MONBUSHO fellow, on leave of absence from ^^ ^^ Al.I. Cuza'' University of Ia\c{s}i, Romania}\, }\thanks{Address after March $1^{\rm st}$ 1996: Dept. of Theoretical Physics, ^^ ^^ Al.I. Cuza'' University, Bd. Copou no.11, 6600 Ia\c{s}i, Romania} \\ \\
{\it Department of Physics} \\ {\it Toyama University}
\\ {\it Gofuku, 930 Toyama, Japan}}
\date{}
\maketitle
\begin{abstract}
Motivated by the recent wave of investigations on plane domain wall spacetimes with 
nontrivial topologies, the present paper deals with (probably) the most simple source 
field configuration which can generate a spatially planary symmetric static spacetime, 
namely a minimally coupled massless scalar field that depends only upon a spacelike 
coordinate, $z$. It is shown that the corresponding exact 
solutions $({\cal M}, {\bf{\rm g}}_{\pm})$ are algebraically special, 
type $D - [S - 3T]_{(11)}$, and represent globally pathologic spacetimes 
with a $G_{4}$ - group of motion acting on ${\bf{\rm R}}^{2} \times {\bf{\rm R}}$ orbits. 
In spite of the model simplicity, these $\phi$ - generated worlds possess naked 
timelike singularities (reached within a finite universal time by normal 
non-spacelike geodesics), are completely free of Cauchy surfaces and contain 
into the $t$ - leveled sections points which can not be jointed 
by ${\rm C}^{1}$ - trajectories images of oblique non-spacelike geodesics. 
Finally, we comment on the possibility of deriving 
from $({\cal M}, {\bf{\rm g}}_{\pm})$ two other physically 
interesting ^^ ^^ $\phi$ - generated'' spacetimes, by appropiate 
jonction conditions in the $(z = 0)$ - plane.
\end{abstract}
\begin{center}
To appear in {\it Foundations of Physics}, {\bf 26} (1996).
\end{center}

\newpage

\section{INTRODUCTION}

It has been done by now a great deal of work on (and within the frame of) the generically 
called Scalar-Tensor Theories of Gravity. Although variously motivated, they lead 
to important results with a large variety of implications in High Energy Physics 
and Cosmology [2, 10, 8, 14, 6]. Nevertheless, General Relativity with its pure 
tensorial gravity and minimally coupled matter (sources) keeps lighting the way to 
an essentially simple understanding of spacetime structure and geometrized gauge 
invariance of basic interactions, strongly supported by fundamental large and small 
scale experiments [15, 17, 21]. In this respect (and especially for highly symmetric 
spacetimes) the Einstein-Klein-Gordon system of the field equations has been 
extensively investigated by both analitical and numerical methods [4, 3]. 
Generally speaking, Inflation and Cosmic Censorship Conjecture have been the main targets. 

Leaving out the coherent scalar fields, for we are not going to deal with in this paper, 
the interest has focussed particularly on bosonic stars, scalar field collapse, 
black holes formation and naked singularities [12, 16, 11, 1]. As a common and very 
important geometro-algebraical feature, the derived spacetimes are spherically 
symmetric for both static and dynamic scalar sources. Although this seems to be a 
natural symmetry (and it actually is for a large class of field sources) the long 
standing investigations on cosmic strings, Bianchi spacetimes, dynamical isotropization 
and more recent on hollow cylinders and topological domain walls, point out that 
the large operating axial or planar symmetries 
might have crucial implications for the Universe [18, 6, 5,19].

Therefore, in this paper we workout the exact solutions of Einstein Equations 
minimally coupled to a static massless scalar field that describe spacetimes 
with a $G_{4}$ - group of motion acting 
on ${\rm{\bf R}}^{2}\times {\rm{\bf R}}$ (2+1)-dimensional orbits. In spite of 
the model simplicity it turns out that the generated spacetimes possess a rather 
complicated structure which involves naked timelike singularities, incompleteness 
with respect to the normal non-spacelike geodesics and the absence of the Cauchy 
surfaces, altogether confirming the recently discovered 
non-trivial topologies of the plane domain wall spacetimes [20].

\section{THE EXACT SOLUTIONS}

As the massless ${\rm{\bf R}}$ - valued field $\phi$ depends only upon a spacelike 
coordinate, an efficient ansatz to get the starting form of the metric is to 
foliate the base manifold ${\cal M} = {\rm{\bf R}}^{4}$ by (2+1) - 
dimensional $\lbrace \phi = {\rm constant} \rbrace $ - surfaces such that the 
privileged (spacelike) coordinate, say $z$, to represent the (universal) 
parameter along the integral curves of the normal $\partial_{\phi}$, i.e.
\begin{equation}
\frac{dz}{d \mu} = 1 = \gamma^{-1} \frac{d \phi}{dz},
\end{equation}
resulting ^^ ^^ ab initio '' the field expression (up to an additive physically unimportant constant)
\begin{equation}
\phi = \gamma z,
\end{equation}
with $\gamma \in {\rm{\bf R}} - \lbrace{0}\rbrace$ a constant of dimensionality. Being interested in planar symmetry and keepping the timelike Killing field $\partial_{t}$ at unity,
\begin{equation}
g \left(\partial_{t} \, , \, \partial_{t} \right) = g_{44} = -1,
\end{equation}
the induced metric on the spacelike $\lbrace z = {\rm const.}, \, t={\rm const}.\rbrace$ sections can be written as
\begin{equation}
(ds^{2})_{\stackrel{{z={\rm const.}}}{\scriptscriptstyle t={\rm const.}}} = e^{2f(z)} \delta_{AB} \, dx^{A} dx^{B} \, ; \; A, B = \overline{1,2} \, ,
\end{equation}
with $x^{1} = x, x^{2} = y$ covering the whole ${\rm{\bf R}}^{2}$ and the (real) function $f \in {\rm C}^{n \geq 2} ({\rm{\bf R}})$ at least piecewisely. Thus, with the norm of $\partial_{z}$ defined by
\begin{equation}
g \left(\partial_{z} \, , \, \partial_{z} \right) = g_{33} = e^{2 h(z)}, \; h \in {\rm C}^{m \geq} ({\rm{\bf R}}),
\end{equation}
the resulting (Lorentzian) metric reads
\begin{equation}
ds^{2} = e^{2f(z)} [(dx)^{2} + (dy)^{2}] + e^{2h(z)} (dz)^{2} - (dt)^{2}
\end{equation}
and the spacetime $({\cal M}, {\rm{\bf g}})$ possesses the Killing vector fields $\partial_{x} \, , \, \partial_{y} \, , \, x \partial_{y} - y \partial_{x} \, , \, \partial_{t}$. 

An interesting observation is that that for the field (2) minimally coupled to (6) the (massless) Gordon equation coincides with the deDonder-Fock gauge fixing condition yielding atonce the local coordinates harmonicity and the ^^ ^^ norm function''
\begin{equation}
h=2f
\end{equation}

With respect to the pseudo-orthonormal frames
\begin{eqnarray}
& a) & e_{1} = e^{-f} \partial_{x}\, , \, e_{2} = e^{-f} \partial_{ y} \, , \, e_{3} = e^{-2f} \partial_{z} \, , \, e_{4} = \partial_{t} \nonumber \\* & & \\*
& b) & \omega^{1} = e^{f} dx \, , \, \omega^{2} = e^{f} dy \, , \, \omega^{3} = e^{2f} dz \, , \, \omega^{4} = dt \nonumber
\end{eqnarray} 
in ${\rm T}({\cal M})$ and ${\rm T}^{*}({\cal M})$ respectively, the curvature has the only essential components
\begin{equation}
R_{1212} = - e^{-4f} (f')^{2} \; ; \; R_{A3B3} = - e^{-4f} [f'' - (f')^{2}] \delta_{AB} ,
\end{equation}
where $(\, \cdot \,)' = \frac{d \cdot}{dz}$ , and consequently the Ricci tensor components are
\begin{equation}
R_{AB} = e^{-4f} f'' \delta_{AB} \, , \; R_{33} = - 2 e^{-4f} [f'' - (f')^{2}]
\end{equation}
casting the field equations (with the minimally coupled massless $\phi$),
\begin{equation}
R_{ab} = \kappa_{0} \phi_{| a} \phi_{| b} \, , \; ({\rm where} \, \phi_{| a} = e_{a} (\phi)),
\end{equation}
into the only algebraically independent and extremely simple differential equation
\begin{equation}
f' = \pm \frac{| \alpha |}{2}
\end{equation}
where we have shifted (for later convenience) $\gamma \rightarrow (2 \kappa_{0})^{-1/2} \cdot \alpha$, such that $[\alpha] = {\rm meter}^{-1}$. Hence, the (disjoint) $(z=0)$ - Minkowskian - like calibrated $({\cal M}, {\rm{\bf g}}_{\pm} , \, \phi)$ - exact solutions are
\begin{eqnarray}
& a) & ds_{\pm}^{2} = exp(\pm | \alpha | z) [(dx)^{2} + (dy)^{2}] + exp(\pm 2 | \alpha | z) \cdot (dz)^{2} - (dt)^{2} \nonumber \\* & & \\*
& b) & \phi(z) = \frac{\alpha z}{\sqrt{2 \kappa_{0}}} \nonumber
\end{eqnarray}  
with the ${\cal M} = {\rm{\bf R}}^{2} \times {\rm{\bf R}} \times {\rm{\bf R}}$ spacetime geometry intrinsically characterized by
\begin{equation}
-R_{1212} = R_{1313} = R_{2323} = \frac{1}{2} R_{33} = \frac{1}{2} R_{33} = \frac{1}{2} R = \frac{\alpha^{2}}{4} exp(\mp 2 | \alpha | z)
\end{equation}
and the source field ^^ ^^ dynamical parameters''
\begin{eqnarray}
& a) & -T_{11} = - T_{22} = T_{33} = T_{44} = w \nonumber \\*
& & \\* & b) & w= \frac{\alpha^{2}}{4 \kappa_{0}} \cdot exp (\mp 2 | \alpha | z), \nonumber
\end{eqnarray}
as measured with respect to (8).

\section{THE ALGEBRAIC CLASSIFICATION}

Since only $R_{33} \neq 0$, the Ricci tensor has two distinct sets of eigenvalues
\begin{eqnarray}
& a) & \lambda_{1} = \lambda_{2} = \lambda_{4} = 0 \nonumber \\* & & \\*
& b) & \lambda_{3} = R \nonumber
\end{eqnarray}
with $\left \lbrace e_{A}, \, e_{4} \right \rbrace_{A=1,2}$ and $ \left \lbrace e_{3} \right \rbrace $ as the corresponding eigenvectors. In what it concerns the essential components of the (conformal) Weyl tensor, they are highly symmetric, being expressed in terms of the scalar curvature alone
\begin{eqnarray}
& a) & C_{1212} = - \frac{R}{3} \nonumber \\*
& b) & C_{A3B3} = \frac{R}{6} \delta_{AB} \\*
& c) & C_{\alpha 4 \beta 4} = \frac{R}{2} \left( \delta_{\alpha 3} \delta_{\beta 3} - \frac{1}{3} \delta_{\alpha \beta} \right), \; \alpha , \beta = \overline{1,3}. \nonumber
\end{eqnarray}
As the right dual ${\tilde{C_{abcd}}} = \frac{1}{2} \varepsilon_{cdef} {C_{ab}}^{ef}$ vanishes identically, the complex Weyl tensor $C_{abcd}^{*}$ coincides with (17) and the (generally complex) (3 x 3) - $Q$ matrix [13]
\begin{equation}
Q_{\alpha \beta} = - C_{\alpha 4 \beta 4}^{*}
\end{equation}
(obtained from $Q_{ab} = - C_{abcd}^{*} u^{c} u^{d} \; {\rm for} \; u=e_{4}$) gets simply
\begin{equation}
Q = {\rm diag} \left( \frac{R}{6} \, , \, \frac{R}{6} \, , \, - \frac{R}{3} \right)
\end{equation}
emphasizing the Weyl eigenbivectors $ \left \lbrace \frac{1}{2} \varepsilon_{A \beta \gamma} \omega^{\beta} \wedge \omega^{\gamma}, \, \omega^{A} \wedge \omega^{4} \right \rbrace_{\beta, \gamma = \overline{1,3}}^{A=1,2}$, for $\mu_{1} = \mu_{2} = \frac{R}{6}$  
and $\lbrace \omega^{1} \wedge \omega^{2}, \omega^{3} \wedge \omega^{4} \rbrace$ for $\mu_{3} = - \frac{R}{3}$. Hence, (16) and (19) point out that, according to the Petrov-Segr\'{e}-Plebanski classification, the ^^ ^^ $\phi$ - generated'' spacetimes $\left({\cal M}, {\rm{\bf g}}_{\pm}\right)$, in (13), are type $D-[S-3T]_{(11)}$ exact solutions that can be hydrodynamically simulated by a tachyonic fluid source with $-p = \frac{1}{3} \rho = w$ and $u^{3} = 1$, i.e.
\begin{equation}
T_{ab}^{(s)} = 2 w \left[ \delta_{a}^{3} \delta_{b}^{3} - \frac{1}{2} \eta_{ab} \right]
\end{equation}

\vspace*{0.5cm}
\section{LOCAL AND GLOBAL IMBEDDINGS}

Since a massless scalar source is much more ^^ ^^ orthodox'' than a tachyonic one and its stress-energy-momentum tensor (15) behaves quite well under all the energy conditions [9], we shall not insist further on this analogy. However, it suggests that something might be special, or at least generically unusual, in the global structure of $\left({\cal M}, {\rm{\bf g}}_{\pm} \right)$. 

For instance, defining on ${\cal M}$ the Ricci-eigenvectors fields
\begin{equation}
X_{(1)}  =  X_{(1)}^{\, A} e_{A} + X_{(1)}^{\, 4} e_{4} \; , \;
X_{(2)}  =  e_{3} 
\end{equation}
and computing the sectional curvature
\begin{equation}
K \left(X_{(1)} , X_{(2)} \right) = \frac{\alpha^{2}}{4} e^{\mp 2 | \alpha | z} \left[ \delta_{CB} X_{(1)}^{\, C} X_{(1)}^{\, B} - \left(X_{(1)}^{\, 4} \right)^{2} \right]^{-1} \cdot \, \delta_{AB} X_{(1)}^{\, A} X_{(1)}^{\, B} ,
\end{equation}
one finds that for a \underline{negative constant} $K$ it always exists an unit timelike vector field $X_{(1)}$ satisfying
\begin{eqnarray}
& a) & \left(X_{(1)}^{\, 1} \right)^{2} + \left( X_{(1)}^{\, 2} \right)^{2} = \frac{4}{\alpha^{2}} (-K) exp(\pm 2 | \alpha | z) \nonumber \\*
& & \\*
& b) & \left(X_{(1)}^{\, 4} \right)^{2} = 1 + \frac{4}{\alpha^{2}} (-K) exp(\pm 2 | \alpha | z) , \nonumber 
\end{eqnarray} 
while for a \underline{positive constant} $K$ the corresponding (unit) spacelike field $X_{(1)}$, having to fulfil
\begin{eqnarray}
& a) & \left(X_{(1)}^{\, 1} \right)^{2} + \left( X_{(1)}^{\, 2} \right)^{2} = \frac{4}{\alpha^{2}} \, K \, exp(\pm 2 | \alpha | z) \nonumber \\*
& & \\*
& b) & \left(X_{(1)}^{\, 4} \right)^{2} = \frac{4}{\alpha^{2}} \, K \, exp(\pm 2 | \alpha | z) - 1 , \nonumber 
\end{eqnarray} 
can be defined only on the ${\rm{\bf R}}^{2} \times \left[ \frac{1}{2 | \alpha |} \ln  
\left(\frac{\alpha^{2}}{4K} \right), \infty \right) \times {\rm{\bf R}}$ region 
of $({\cal M}, {\rm{\bf g}}_{+})$ and ${\rm{\bf R}}^{2} \times 
\left( \infty, - \frac{1}{2 | \alpha |} \ln  \left(\frac{\alpha^{2}}{4K}\right) \right] 
\times {\rm{\bf R}}$ region of $({\cal M}, {\rm{\bf g}}_{-})$ respectively. 
This is related to the fact that, setting for example  $K=(\alpha/2)^{2}$, the corresponding ^^ ^^ bordered'' spacetimes $\left({\rm{\bf R}}^{2} \times {\rm{\bf R}}_{\pm} \times {\rm{\bf R}} \, , \, {\rm{\bf g}}_{\pm} \right)$ can be imbedded as the 4-surfaces
\begin{eqnarray}
X & = & x \, e^{\pm \frac{| \alpha | z}{2}} \nonumber \\*
Y & = & y \, e^{\pm \frac{| \alpha | z}{2}} \nonumber \\*
Z & = & \pm (3 | \alpha |)^{-1} \left \lbrace \left[ e^{\pm | \alpha | z} - \frac{3}{4} \alpha^{2} (x^{2} + y^{2}) + 3 \right] e^{\pm \frac{| \alpha | z}{2} } - 4 \right \rbrace  \\*
W & = & \pm (3 | \alpha |)^{-1} \left \lbrace \left[e^{\pm | \alpha | z} - \frac{3}{4} \alpha^{2} (x^{2} + y^{2}) - 3 \right] e^{\pm \frac{| \alpha | z}{2}} + 2 \right \rbrace \nonumber \\*
T & = & t \nonumber
\end{eqnarray}
in the (abstract) 5-dimensional Euclidean space ${\rm{\bf R}}^{5}$ of signature 1,
\begin{equation}
dS_{(1)}^{\, 2} = (dX)^{2} + (dY)^{2} + (dZ)^{2} - (dT)^{2} - (dW)^{2},
\end{equation}
but this doesn't work for their ^^ ^^ $z$ - complements'' (${\rm{\bf R}}^{2} \times {\rm{\bf R}}_{\mp} \times {\rm{\bf R}} \, , \, {\rm{\bf g}}_{\pm}$) since (25) ceased to be an one-to-one immersion. Therefore, a higher dimensional Euclidean space is needed for imbedding the ^^ ^^ $\phi$ - generated Worlds'' (13). This is offered by ${\rm{\bf R}}^{6}$ endowed with the metric (of 0 - signature)
\begin{equation}
dS_{(0)}^{\, 2} = (dX^{1})^{2} - (dX^{2})^{2} + (dY^{1})^{2} - (dY^{2})^{2} + (dZ)^{2} - (dT)^{2}
\end{equation}
wherein the spacetimes $({\cal M} , {\rm{\bf g}}_{\pm})$ appear respectively as positive $(X^{2} > 0, \, Y^{2} > 0, \, Z > 0)$, and negative $(X^{2} > 0, \, Y^{2} > 0, \, Z < 0)$, wings of an exponential (4-dimensional) hyperboloid, i.e.
\newpage
\begin{eqnarray}
& a) & X^{1} = \frac{2}{| \alpha |} e^{\pm \frac{ |
 \alpha |}{2} z} \sinh \left(\frac{| \alpha |}{2} x \right) \; ; \;  X^{2} = \frac{2}{| \alpha |} e^{\pm \frac{| \alpha |}{2} z}  \cosh \left(\frac{| \alpha |}{2} x \right) \nonumber \\*
& b) & Y^{1} = \frac{2}{| \alpha |} e^{\pm \frac{| \alpha |}{2} z} \sinh \left(\frac{| \alpha |}{2} y \right) \; ; \; Y^{2} = \frac{2}{| \alpha |} e^{\pm \frac{| \alpha |}{2} z} \cosh \left(\frac{| \alpha |}{2} y \right) \\*
& c) & Z=\pm \frac{1}{| \alpha |} \lbrace [e^{\pm | \alpha | z}+1]^{1/2} \cdot e^{\pm \frac{| \alpha |}{2} z} +  \ln [(e^{\pm | \alpha | z} + 1)^{1/2} + e^{\pm \frac{| \alpha |}{2} z} ] \rbrace \nonumber \\*
& d) & T=t \nonumber
\end{eqnarray}
\marginpar{figs. 1, 2}
\\
As it can be noticed inspecting figs. 1 and 2, that show the $t$ - leveled 2-dimensional 
hyperboloidal wings (28.$a,c$) in the conformal $z$ - 
reparametrization $exp \left(\pm \frac{| \alpha |}{2} z \right) 
\rightarrow {\underline{z}} \in {\rm{\bf R}}_{+}$, 
the $\lbrace x, {\underline{z}} \rbrace$ - system of congruences covers the 
two surfaces except for the timelike singularity ${\underline{z}}=0$ 
(located in $X^{1} = X^{2} = Y^{1} = Y^{2} = Z = 0$) that corresponds to 
the (2+1) - surfaces of transitivity $\lbrace z= - \infty \rbrace$ 
and $\lbrace z= \infty \rbrace$ in $({\cal M}, {\rm{\bf g}}_{\pm})$ respectively. 
Nevertheless, the ($x$ = constant) - trajectories are boldly reaching the 
pointwise projection of these two singularities which is a clear evidence 
of their nakedness with respect to the normal non-spacelike geodesics. 
Another intriguing feature possessed by the exact solutions (13) is that, 
as it is shown on the positive wing in fig. 2, there is no oblique non-spacelike 
geodesic that touches the singularity and moreover, the wing is just partially 
covered by any family of oblique trajectories starting from their turning point. 
Hence, each of the worlds $\left({\cal M} , {\rm{\bf g}}_{\pm} \right)$ contains 
in its ($t$ = const.) leaves points which can not be joined by ${\rm C}^{1}$ - trajectories, 
images of the oblique non-spacelike geodesics. As it has been stressed by Geroch [7], 
this is an undeniable sign of a globally pathologic spacetime structure. 

\section{PENROSE DIAGRAMS. CONCLUSIONS}

Since $({\cal M} , {\rm{\bf g}}_{-})$ can be formally obtained from $({\cal M} , {\rm{\bf g}}_{+})$ by the parity transformation $z \rightarrow -z$, the Penrose diagram of the  $({\cal M} , {\rm{\bf g}}_{+})$ - world is quite sufficient. With the conformal $z$ - reparametrization 
\begin{equation}
\underline{z} = \alpha^{-1} exp(\alpha z)
\end{equation}
we get an ${\rm{\bf R}}^{2} \times {\rm{\bf R}}_{+} \times {\rm{\bf R}}$ topology of the base manifold ${\cal M}$ and the induced metric
\begin{equation}
ds^{2} = \alpha \underline{z} \delta_{AB} \, dx^{A} dx^{B} + ( d \underline{z})^{2} - (dt)^{2}
\end{equation}
suggests the definition of the advanced, respectively retarded, Penrose null-coordinates 
\begin{equation}
u = \arctan(t+ \underline{z}) \; , \; v = \arctan(t - \underline{z})
\end{equation}
Because of $\underline{z} \geq 0$, the $\phi$ - generated world covers just a half of the ^^ ^^ Penrose square'' 
$\left[- \frac{\pi}{2} \, , \, \frac{\pi}{2} \right] \times \left[- \frac{\pi}{2} \, , \, \frac{\pi}{2} \right]$, namely the one given by
\begin{equation}
u-v \geq 0
\end{equation}
\marginpar{fig. 3}
\\
As fig. 3 shows, there can be noticed the following abnormal large scale properties. 
The $\underline{z} = 0$ singularity is a timelike one and 
consequently $({\cal M}\, , \, {\rm{\bf g}}_{+})$ does not possess 
neither $z = - \infty$ past and future null infinities, ${\cal I}^{-}(z = -\infty)$ 
and ${\cal I}^{+}(z = -\infty)$, nor Cauchy surfaces, and is incomplete with 
respect to the normal non-spacelike geodesics. The oblique or normal timelike 
geodetic observers moving \underline{towards $z= \infty$} reach $i^{+}$ and so 
they have no future event horizons. However, the past extensions of the normal 
timelike geodesics reach the singularity within a finite universal time and 
simply disappear from $({\cal M}\, , \, {\rm{\bf g}}_{+})$. In other words, 
when such an observer appears through the naked singularity $\underline{z} = 0$, 
he has a past event horizon because of his future light cone creation. 
Similarly, there are future event horizons for the normal timelike geodetic 
observers moving \underline{towards $z = - \infty$} because of their past 
light cone destruction at the moment of disappearance through the singularity. 
However, their past extended world-lines reach $i^{-}$ together with the rest 
of the oblique timelike geodesics and consequently there are no past event 
horizons for them. Doubtless, it does exist a strange and complicated structure 
of the two $\phi$ - generated worlds which strongly affects the non-spacelike motions, 
spoils the Cauchy predictments and violates the Cosmic Censorship Conjecture. 
Nevertheless, it seems interesting to observe that, inspired by the sectional 
curvature problem discussed in Sect. 4, one might think of the two qualitatively 
utterly different spacetimes:  \[ds_{1}^{2} = exp(- \alpha | z |) [(dx)^{2} + (dy)^{2}] + exp(- 2 \alpha | z |) \cdot (dz)^{2} - (dt)^{2} \] and 
\[ ds_{2}^{2} = exp(\alpha | z |) [(dx)^{2} + (dy)^{2}] + exp(2 \alpha | z |) \cdot (dz)^{2} - (dt)^{2} \, , \]
with $\alpha > 0$ and the base manifold ${\cal M} = {\bf{\rm R}}^{2} \times {\bf{\rm R}}$, 
that can be obtained from $({\cal M} , {\bf{\rm g}}_{\pm})$ by suitable jonction conditions 
on the $\lbrace z=0 \rbrace$ - hypersurface. The former, $({\cal M} , {\bf{\rm g}}_{1})$, 
will be highly singular on its asimptotic regions (possessing timelike curvature 
singularities at both $z = \mp \infty$), while the latter, $({\cal M} , {\bf{\rm g}}_{2})$, 
provides a non-trivial example of an asimptotically flat spacetime, with ^^ ^^ $\delta$ - like'' curvature singularities onto the plane of jonction. 

However, for stating their physical consistency, further work is required in order to clarify the nature and the (probably discrete) distribution of the additional sources into the $(z=0)$ - plane, which are seemingly necessary for ^^ ^^ supporting'' both the connection and curvature discontinuities.        
    
\vspace{1cm}{\large{\bf ACKNOWLEDGMENTS}} \\ \\
It is a pleasure to thank Shinji Hamamoto and Misao Sasaki for fruitful discussions on an early version of the paper. The financial support from the Japanese Government under a Monbusho Felloship is gratefully acknowledged.

\begin{figure}[p]
\vspace*{10cm}
\caption{A general 3D - view of the worlds $({\cal M}, {\bf{\rm g}}_{\pm})$ at constant universal time, with the coordinates $Y^{1}$ and $Y^{2}$ omitted. On the $Z$ - axis the imbedding has been calibrated such that the point $X^{1} = X^{2} = Z = 0$ represents the projection of the naked singularity $z = - \infty$ and $z = + \infty$ of the worlds $({\cal M}, {\bf{\rm g}}_{+})$ and $({\cal M}, {\bf{\rm g}}_{-})$, respectively.}
\end{figure}
\begin{figure}[p]
\vspace*{10cm}
\caption{Same as in Fig.1 but from a frontal view point. One can notice on the positive wing the family of $y = const.$ trajectories projecting the oblique null geodesics strating from $x = 0, z = z_{0}$. None of them is touching the singularity and the family does not cover the wing completely.}                                                                                                                                                                                
\end{figure}

\begin{figure}[p]
\hspace*{3cm} $i^{+}$ \\
\vspace*{6.5cm} \\
\hspace*{10.5cm} $i^{0}$ \\
\vspace*{6.5cm} \\
\hspace*{3cm} $i^{-}$
\caption{Penrose diagram of the $({\cal M}, {\bf{\rm g}}_{+})$ - world. The bold line represents the physical (timelike) singularity, dashed lines are the normal null geodesics and the solid curves represent the maximally extended (into the $({\cal M}, {\bf{\rm g}}_{+})$ spacetime) normal timelike geodesics of the observers strating (at $t=0$) from different $z$ = constant planes. }
\end{figure}
\end{document}